\documentclass[11pt,a4paper]{article}

\usepackage{amsmath,amsfonts,amsbsy,amssymb,array,accents,dsfont}
\usepackage{enumerate,array,latexsym,graphicx,mathrsfs,verbatim,psfrag}
\usepackage{bm} 
\usepackage{bbm,braket}
\usepackage[normalem]{ulem}
\usepackage{booktabs} 
\usepackage[usenames]{color}

\usepackage[UKenglish]{babel}
\usepackage{fancybox}

\usepackage[nosort]{cite}
\usepackage{chngpage} 

\usepackage{enumitem}
\setlist{itemsep=0pt}

\usepackage[colorlinks=true,      linkcolor=darkblue,      urlcolor=darkblue,      
            filecolor=darkblue,      citecolor=darkblue,       pdfstartview=FitH,     
						pdfpagemode=UseNone,      bookmarksopen=true]{hyperref}  
\usepackage[all]{hypcap}     

\newcommand{\captionfonts}{\small}
\makeatletter  
\long\def\@makecaption#1#2{%
  \vskip\abovecaptionskip
  \sbox\@tempboxa{{\captionfonts #1: #2}}%
 \ifdim \wd\@tempboxa >\hsize
    {\captionfonts #1: #2\par}
  \else
    \hbox to\hsize{\hfil\box\@tempboxa\hfil}%
  \fi
  \vskip\belowcaptionskip}
\makeatother   

\DeclareMathSymbol{\medhatsym}{\mathord}{largesymbols}{"62} 
\newcommand\lowermedhatsym{
  \text{\smash{\raisebox{-1.28ex}{%
    $\medhatsym$}}}}
\newcommand\medhat[1]{
  \mathchoice
    {\accentset{\displaystyle\lowermedhatsym}{#1}}
    {\accentset{\textstyle\lowermedhatsym}{#1}}
    {\accentset{\scriptstyle\lowermedhatsym}{#1}}
    {\accentset{\scriptscriptstyle\lowermedhatsym}{#1}}
}

\DeclareMathSymbol{\medtildesym}{\mathord}{largesymbols}{"65}
\newcommand\lowermedtildesym{
  \text{\smash{\raisebox{-1.35ex}{%
    $\medtildesym$}}}}
\newcommand\medtilde[1]{
  \mathchoice
    {\accentset{\displaystyle\lowermedtildesym}{#1}}
    {\accentset{\textstyle\lowermedtildesym}{#1}}
    {\accentset{\scriptstyle\lowermedtildesym}{#1}}
    {\accentset{\scriptscriptstyle\lowermedtildesym}{#1}}
}

\def\({\left(}
\def\){\right)}
\def\[{\left[}
\def\]{\right]}

\def\barray{\begin{array}}
\def\earray{\end{array}}
\def\be{\begin{equation}}
\def\ee{\end{equation}}
\def\bea{\begin{eqnarray}}
\def\eea{\end{eqnarray}}
\def\bal{\begin{align}}
\def\eal{\end{align}}

\mathchardef\mhyphen="2D

\newcommand{\tr}{\mathrm{Tr}\;\!}

\numberwithin{equation}{section} %

\makeatletter
\g@addto@macro\bfseries{\boldmath}
\makeatother

\definecolor{cardinal}{rgb}{0.6,0,0}
\definecolor{darkgreen}{rgb}{0,0.4,0}
\definecolor{purple}{rgb}{0.5, 0, 0.5}
\definecolor{golden}{rgb}{0.92, 0.7, 0}
\definecolor{midnight}{rgb}{0, 0, 0.5}
\definecolor{darkblue}{rgb}{0, 0, 0.8}

\definecolor{emeraude}{RGB}{34,120,15}
\definecolor{turquoise}{RGB}{49, 140, 231}
\definecolor{framboise}{RGB}{199, 44, 72}

\def\cC{{\cal C}}

\def\cN{{\cal N}}
\def\cO{{\cal O}}

\def\cT{{\cal T}}

\newcommand{\T}[3]{\ensuremath{ #1{}^{#2}_{\phantom{#2} \! #3}}}		

\begin{document}

\begin{flushright}
%
%
\end{flushright}

\vspace{18mm}

\begin{center}

{\huge \bf{Precision holography of \vspace{5mm}\\ AdS$_5$ bubbling geometries}}

\vspace{22mm}

{\large
\textsc{David Turton,\;\,Alexander Tyukov}}

\vspace{15mm}

\baselineskip=16pt
\parskip=3pt

Mathematical Sciences and STAG Research Centre,\\ University of Southampton,\\ Highfield,
Southampton SO17 1BJ, UK \\

\vspace{12mm}

{\upshape\ttfamily d.j.turton @ soton.ac.uk, a.tyukov @ soton.ac.uk} \\

\vspace{23mm}

\textsc{Abstract}

\end{center}

\begin{adjustwidth}{10mm}{10mm} 
%
\vspace{3mm}
\baselineskip=14.5pt
\noindent
Holographic duality provides a microscopic interpretation of asymptotically Anti-de Sitter supergravity solutions. 
The dual states of the field theory give rise to expectation values of light operators.
These expectation values correspond to coefficients in the asymptotic expansion of gauge-invariant combinations of supergravity fields. 
We consider the duality between type IIB theory on AdS$_5 \times $S$^5$ and 4D $\mathcal{N}=4$ super Yang-Mills theory, focusing on the half-BPS sector. 
We clarify the structure of the precision holographic dictionary, making explicit the distinction between single-trace operators and single-particle operators, where the latter contain admixtures of multi-traces. 
We rewrite the holographic dictionary for half-BPS operators of dimension four in the single-particle basis.
We then apply this dictionary to perform precision holographic studies of two different smooth supergravity solutions in the class derived by Lin, Lunin and Maldacena, that have been recently used to compute all-light four-point correlators by making extrapolations of heavy-light correlators.

\end{adjustwidth}

\thispagestyle{empty}

\newpage


%
%


\baselineskip=16pt
\parskip=3pt


\section{Introduction}

Holographic duality is the remarkable conjecture that a gravitational theory in $d+1$ dimensions (possibly times other compact directions) may be completely equivalent to a non-gravitational theory that lives on the $d$-dimensional boundary of the original space-time~\cite{Maldacena:1997re}. Most typically, the gravitational theory lives in $(d+1)$-dimensional Anti-de Sitter space (AdS$_{d+1}$), and the non-gravitational theory is a conformal field theory (CFT$_d$).

On the gravitational side of such a duality, there can be large families of asymptotically Anti-de Sitter supergravity solutions that are horizonless and smooth up to possible physical sources. Holography relates these solutions to generically heavy pure states of the dual field theory.
An interesting class of such solutions are those that are backreacted coherent bound states of many supergravitons, see for example~\cite{Lunin:2001jy,Lin:2004nb,Bena:2015bea,Bena:2016ypk,Bena:2017xbt,Ceplak:2018pws,Heidmann:2019xrd,Ceplak:2024dbj}.
For many of these families of supergravity solutions, there is a detailed proposal for the dual CFT states.

Several of these proposed identifications between supergravity solutions and CFT states have passed non-trivial tests using precision holographic studies of protected heavy-light three-point correlation functions~\cite{Giusto:2015dfa,Giusto:2019qig,Rawash:2021pik,Ganchev:2021ewa}, using the formalism developed in~\cite{Skenderis:2006uy,Kanitscheider:2006zf,Skenderis:2007yb}.
To perform such studies, one constructs gauge-invariant combinations of supergravity fields, and studies their asymptotic expansion at large radial distance in AdS.
Coefficients in this expansion are dual to expectation values of light operators in the corresponding heavy CFT state.

Smooth supergravity solutions have also been used to compute heavy-light and all-light correlators, including very recently~\cite{Turton:2024afd,Aprile:2024lwy,Aprile:2025hlt} in the duality between type IIB theory on AdS$_5 \times$ S$^5$ and 4D $\mathcal{N}=4$ SU($N$) super Yang-Mills (SYM) theory~\cite{Maldacena:1997re}. These works used half-BPS asymptotically AdS$_5 \times$S$^5$ smooth horizonless supergravity solutions in the class derived by Lin, Lunin and Maldacena (LLM)~\cite{Lin:2004nb}. The correlators were computed in the supergravity regime in the bulk, i.e.~at large $N$ and large 't Hooft coupling.

These recent AdS$_5$/CFT$_4$ developments build on similar work in AdS$_3$/CFT$_2$ duality on heavy-light four-point correlators~\cite{Galliani:2017jlg,Bombini:2017sge,Bombini:2019vnc,Giusto:2023awo} and computing all-light correlators by making extrapolations thereof~\cite{Giusto:2018ovt,Giusto:2019pxc,Giusto:2020neo,Ceplak:2021wzz}; see also the related works~\cite{Bena:2019azk,Rastelli:2019gtj,Giusto:2020mup}.
Here, `heavy' denotes CFT operators whose conformal dimensions scale linearly with the central charge $c$ in the large $c$ limit, and `light' denotes CFT operators whose conformal dimensions are independent of $c$ in the large $c$ limit.

The recent computations of four-point correlators in~\cite{Turton:2024afd,Aprile:2024lwy,Aprile:2025hlt} used two different LLM geometries. 
Each of these solutions is defined by a profile function that describes the boundary of a single droplet in the free fermion description of the LLM solutions.
The work of~\cite{Turton:2024afd} used a profile function that involves a single mode, that had been studied in~\cite{Skenderis:2007yb}. 
By contrast, the work of~\cite{Aprile:2024lwy,Aprile:2025hlt} used a solution~\cite{Giusto:2024trt} that lies in the consistent truncation of~\cite{Cvetic:2000nc}, that was first constructed and studied in~\cite{Chong:2004ce,Liu:2007xj}, and corresponds to an elliptical profile function~\cite{Chen:2007du}.
For a selection of other recent work involving LLM solutions, see~\cite{Balasubramanian:2018yjq,Berenstein:2023vtd,Eleftheriou:2023jxr,Deddo:2024liu,Chen:2024oqv}.

The primary motivations for this work are to make precise the difference between the two LLM backgrounds studied in~\cite{Turton:2024afd,Aprile:2024lwy,Aprile:2025hlt}, and to investigate the holographic description of these solutions with AdS$_5$/CFT$_4$ precision holography~\cite{Skenderis:2006uy,Skenderis:2007yb}.

In this paper we study these two solutions perturbatively in a small parameter $\alpha$. The backgrounds agree to order $\alpha$, but differ at order $\alpha^2$. 
At first order in $\alpha$, the linearised supergravity fields are among the fluctuations around global AdS$_5\times $S$^5$ classified in~\cite{Kim:1985ez}, as shown in~\cite{Grant:2005qc}.
These fields are dual to 
the state $\ket{O_2}$ corresponding to 
the chiral primary operator $O_2\sim\tr Z^2$, where $Z$ is a holomorphic combination of two of the six hermitian adjoint scalars of the dual field theory, $Z = \frac{1}{\sqrt{2}}(\Phi^1 + i \Phi^2)$.

Beyond the linearised order, in general LLM solutions are dual to multi-traces built from powers of the field $Z$. These operators have scaling dimension equal to a particular U(1) R-charge in the SO(6) R-symmetry group, $\Delta=J$.
In~\cite{Giusto:2024trt} it was proposed that the LLM solution in the consistent truncation is dual to a coherent state composed only of powers of $O_2$. This proposal was refined very recently  in~\cite{Aprile:2025hlt}. We will find supporting evidence for this proposal.
We shall furthermore demonstrate that, by contrast, the solution defined by the single-mode profile is dual to a state that contains the dimension-four chiral primary $O_4$ at order $\alpha^2$, and fix the coefficient of $O_4$ in the state at this order.

To establish these results, we revisit the AdS$_5$/CFT$_4$ precision holographic dictionary, and first clarify that it was originally derived in the single-trace basis of the dual CFT~\cite{Skenderis:2006uy,Skenderis:2007yb}.
We then rewrite the dictionary in the single-particle basis, which will be much more convenient for our analysis. 
Single-particle CFT operators are defined to be half-BPS operators that are orthogonal to all multi-trace operators~\cite{Aprile:2018efk,Aprile:2020uxk}.
They are proposed to be dual to single-particle supergravity states on global AdS$_5 \times $S$^5$.
This generalizes the earlier work of~\cite{Arutyunov:1999en,Arutyunov:2000ima,Arutyunov:2018neq}.
In the single-particle basis, the holographic dictionary takes a simpler form in terms of the 5D supergravity fields that arise from Kaluza-Klein reduction.
The analogous single-particle basis has also been recently used in AdS$_3$/CFT$_2$ holography~\cite{Rawash:2021pik}.

We then use the precision holographic dictionary in the single-particle basis to study the two LLM geometries in question. To do so, on the supergravity side, we convert each background into de Donder-Lorentz gauge, building on the results of~\cite{Turton:2024afd}. On the CFT side, we make an Ansatz for the dual CFT states up to order $\alpha^2$.

We compute expectation values of the single-particle chiral primary operators of dimension two and four, and of certain SO(6) R-symmetry descendants thereof.
The first few expectation values we compute are used to fix the coefficients in the CFT states, thus distinguishing the dual CFT states to the respective LLM backgrounds.
The remaining expectation values, in particular those of the R-symmetry descendants, represent non-trivial cross-checks of the dual CFT states, and also of the holographic dictionary in the single-particle basis.
These cross-checks include an operator whose expectation value arises at order $\alpha^3$.

The structure of this paper is as follows. In Section \ref{sec:holo-ads5} we present the holographic dictionary in the single-particle basis, as well as the explicit form of the light operators whose expectation values we compute, including a set of R-symmetry descendants of the single-particle chiral primaries.
In Section \ref{sec:backgd-fields} we describe the two LLM solutions we consider, and make explicit the difference between the solutions in supergravity.
In Section \ref{sec:evs-alpha-2} we analyze the full set of holographic expectation values up to order $\alpha^2$, and fix the dual CFT states to the two LLM solutions.
In Section \ref{sec:evs-alpha-3} we perform a non-trivial cross-check of the heavy state dual to the solution defined by the single-mode profile by computing an expectation value at order $\alpha^3$. 
In Section~\ref{sec:disc} we discuss our results.


\section{Precision AdS$_5$/CFT$_4$ holography in the single-particle basis}
\label{sec:holo-ads5}

In this section we start by presenting the light operators whose expectation values we compute in this paper. We then discuss the relation between extremal three-point correlators and single-particle operators. We review the relevant parts of the precision holographic dictionary in AdS$_5$~\cite{Skenderis:2006uy,Skenderis:2007yb}, and clarify that the dictionary was derived in the trace basis. We then rewrite the dictionary in the single-particle basis.

\subsection{Single-particle chiral primaries}

Single-particle operators are defined to be orthogonal to all multi-trace operators, i.e.~to have vanishing two-point functions with all such operators~\cite{Aprile:2018efk,Aprile:2020uxk}. We work with gauge group SU($N$) and with complex combinations of the six hermitian scalar fields $\Phi^i$ as follows,
\be
    Z(x)\,=\, \frac{1}{\sqrt{2}}\left(\Phi^1(x)+i\Phi^2(x)\right)\,, \qquad \bar{Z}(x)\,=\, \frac{1}{\sqrt{2}}\left(\Phi^1(x)-i\Phi^2(x)\right)\,, 
\ee
and similarly for $X,\bar{X},Y,\bar{Y}\,$ in terms of $\Phi^a$, where $a$ runs over $3,\ldots,6$.

We absorb the appropriate factors of the Yang-Mills coupling and of $2\pi$ into the definition of the operators, as is often done (see e.g.~\cite{Corley:2001zk,Brown:2007bb,Skenderis:2007yb}), such that the two-point correlators take the form 
\be
\label{eq:2pt-fn-scalars}
    \langle \T{\bar{Z}}{p}{q}(x) \T{Z}{r}{s}(y) \rangle \,=\,
    \frac{\T{\delta}{p}{s} \T{\delta}{r}{q} - \frac1N \T{\delta}{p}{q} \:\! \T{\delta}{r}{s}
    }
    {|x-y|^2} \,.
\ee
We shall work at leading order in large $N$. The $1/N$ term in the propagator is subleading (see e.g.~\cite{Aharony:1999ti}) and shall play no role in the present work. 
From now onwards, we shall suppress the spacetime dependence in most equations.

We first introduce the following notation for non-unit-normalized single-trace chiral primary operators, 
\be
\label{eq:non-normalised-Tn-defn}
\medhat{T}_k \;=\; \tr(Z^k)\;.
\ee
As usual, it is convenient to define single-trace operators that are unit-normalized at leading order in large $N$, with $\Delta=J=k$, see e.g.~\cite{Lee:1998bxa},
\be
\label{eq:normalizedTn}
T_k \;=\; \frac{\tr(Z^k)}{\sqrt{k}\:\! N^{\frac{k}{2}}} \;.
\ee

The unit-normalized single-particle chiral primaries take the following form at leading order in large $N$~\cite{Aprile:2020uxk},
\begin{align}
\label{eq:normalized-On}
  O_2 \;&=\; 
  T_2 \;,
  \qquad
  O_4 \;=\; 
  T_4 - \frac{2}{N} 
  (O_2)^2 \;,
\end{align}
where we have kept only the leading term in the coefficient of the double-trace $(O_2)^2$.
We emphasize that this term can contribute at leading order in large $N$ in extremal and heavy-light correlators.

\subsection{Single-particle R-symmetry descendants}

In this section we work out the explicit form of the single-particle R-symmetry descendants whose expectation values we shall compute in this paper. 

First, we recall that the states dual to LLM solutions involve multi-traces composed of powers of the field $Z$. Such states break the SO(6) R-symmetry to SO(4). 
We will compute expectation values of some SO(4)-invariant R-symmetry descendants of chiral primaries.
We thus define $T_{k,m}$ to be the SO(4) invariant R-symmetry descendants of the single-trace operators $T_k$ that have charge $m$ under the U(1) selected by $Z$, and that are unit-normalized. 
Likewise, we define $O_{k,m}$ to be the analogous descendants of the single-particle chiral primaries $O_k$. We work to leading order in large $N$ throughout.

We start with the single-trace operators. The explicit forms of the descendants of single-trace CPOs can be obtained directly from the relevant spherical harmonic in terms of the six scalars $\Phi^i$~\cite{Lee:1998bxa} (see below around Eq.~\eqref{eq:t-c}). 
Alternatively, one can act on the chiral primary with the 
following SO(4)-invariant lowering operator~\cite{Giusto:2024trt}, which lowers the U(1) $R$-charge by 2 units: 
\be
   J^{-} \,\equiv\, \sum_{a=3}^{6}
    J^{-a}\:\! J^{-a} \;, \qquad~~
    J^{-a} \,\equiv\, (J^{1a}-i J^{2a})\;, \qquad a=3,\ldots,6,
\ee
where 
\be
    J^{ij} \,=\, -i \, \tr \left(\Phi^{i} \frac{\partial}{\partial \Phi^j}-\Phi^{j} \frac{\partial}{\partial \Phi^i}\right) \,, \qquad~~~ i\,,j = 1,\ldots,6\,.  
\ee
For instance, the neutral descendant $O_{2,0}=T_{2,0}$ is
\be
\label{eq:O20}
    O_{2,0}  
    \,=\,
    \frac{1}{\sqrt{6}N} \mathrm{Tr}\Big(2 Z \bar{Z} -\frac12 \Phi^a\Phi^a \Big)
    \,=\,
    \frac{1}{\sqrt{6}N} \mathrm{Tr}\Big(2 Z \bar{Z} -(X\bar{X} + Y\bar{Y}) \Big)
    \,.
\ee
At dimension four, the single-trace charge-two descendant
$T_{4,2}\sim J^{-}T_4$, takes the form
\begin{equation}
    T_{4,2} \,=\, \frac{1}{\sqrt{10}N^2} \mathrm{Tr}\Big(2 Z^3 \bar{Z} -Z^2\Phi^a\Phi^a- \frac12 Z \Phi^a Z \Phi^a \Big)\,,  
\end{equation}
and the single-trace dimension-four neutral descendant $T_{4,0}$ is
\begin{align}
\label{eq:T40}
    T_{4,0}  \,=\, \frac{1}{2\sqrt{5}N^2} \mathrm{Tr}  \bigg[ 
    &
    2 Z^2 \bar{Z}^2 + (Z\bar{Z})^2 
    -\Big(    (Z\bar{Z}+\bar{Z}Z)\Phi^a\Phi^a     +Z\Phi^a\bar{Z}\Phi^a  \Big)
     \\
    & \!
    + \frac16 
    \Big(
    \Phi^a\Phi^a
    \Phi^b\Phi^b
    + \frac12 \Phi^a\Phi^b \Phi^a\Phi^b \Big) \bigg]
  \,.
  \nonumber
\end{align}

We next discuss the single-particle descendants $O_{4,2}$ and $O_{4,0}$. 
To obtain their explicit expressions, one can either act with the above SO(4)-invariant lowering operator on $O_4$, or one can add all possible SO(4)-invariant combinations of double-traces to $T_{4,2}$ and $T_{4,0}$ and fix their coefficients by imposing orthogonality with all double-traces. We obtain
\be
    O_{4,2} \,=\, T_{4,2} - \frac{4\sqrt{3}}{\sqrt{10}N} O_2 O_{2,0} + \frac{2}{\sqrt{10}N^3} \mathrm{Tr} Z \Phi^a \, \mathrm{Tr} Z \Phi^a  
\ee
and
\begin{align}
     O_{4,0} \,&=\, T_{4,0} - \frac{3}{\sqrt{5}N} \left(O_{2,0}\right)^2 -\frac{2}{\sqrt{5}N} O_2^{\dagger} O_2 + \frac{2}{\sqrt{5}N^3} \mathrm{Tr} Z \Phi^a \, \mathrm{Tr} \bar{Z} \Phi^a \cr & \qquad\qquad\quad
     - \frac{1}{6\sqrt{5}N^3}\left( \mathrm{Tr} \Phi^a \Phi^b \, \mathrm{Tr} \Phi^a \Phi^b - \frac{1}{4}\left( \mathrm{Tr} \Phi^a \Phi^a \right)^2\right)  \,. 
\end{align}
In the above expressions, only the terms that are composed solely of $Z$, $\bar{Z}$ contribute to the expectation values of these operators in LLM geometries. We have included these terms for completeness, and for possible use in future work. However, for the applications in Sections \ref{sec:evs-alpha-2} and \ref{sec:evs-alpha-3}, all terms that involve any fields other than $Z,\bar{Z}$ can be ignored.

\subsection{Extremal three-point functions and single-particle operators}

More generally, let us consider the following single-trace operators,
\begin{align}
\label{eq:t-c}
\widehat{T}_{I} \,&=\, C^I_{i_1 \cdots i_k} 
\tr (\Phi^{i_1}\cdots \Phi^{i_k}) \,,
\end{align}
where $C^I$ is a totally symmetric traceless rank-$k$ tensor of SO(6). These operators live in half-BPS multiplets~\cite{Lee:1998bxa}. For details and conventions, see Appendix~\ref{app:conventions}. 
For our holographic applications, we focus on operators up to dimension four, in the SU($N$) theory. Thus the mixing in the single-particle operators involves at most double-trace operators, and no triple or higher traces.

Let us review the relation between extremal three-point functions and the mixing coefficients of double-traces in single-particle operators~\cite{Arutyunov:1999en,Arutyunov:2000ima,Arutyunov:2018neq}.
Consider three unit-normalized single-trace operators
$T_{I_1}$, $T_{I_2}$ and $T_{I_3}$, such that $k_1=k_2+k_3$. The protected extremal three-point function takes the form~\cite{Lee:1998bxa}
\be
    \langle T_{I_1}(x_1) T_{I_2}(x_2) T_{I_3}(x_3) \rangle \,=\, \frac{1}{N} \frac{\sqrt{(k_2+k_3) k_2 k_3} \, \langle C^{I_1} C^{I_2} C^{I_3} \rangle}{|x_1-x_2|^{2k_2} |x_1-x_3|^{2k_3}} \;, 
\ee
where $\langle C^{I_1} C^{I_2} C^{I_3} \rangle$ is the unique SO(6) invariant formed by contracting all indices in the tensors $C^{I_a}$, Eq.~\eqref{eq:trip-c},
and is related to the triple intersection of spherical harmonics $a_{I_1 I_2 I_3}$, Eq.~\eqref{eq:trip}, by 
\be
\label{eq:a-c-main}
     a_{I_1 I_2 I_3}\,=\, \frac{1}{\left(\frac{1}{2}\Sigma+2\right)! \, 2^{\frac{1}{2}\left(\Sigma-2\right)}}\frac{k_1! k_2! k_3!}{\alpha_1! \alpha_2!\alpha_3!}\langle C^{I_1} C^{I_2} C^{I_3} \rangle \,,
\ee
where $\Sigma=k_1+k_2+k_3$, $\alpha_1=\frac12(k_2+k_3-k_1)$, and similarly for $\alpha_2,\alpha_3$. At extremality, i.e.~for $k_1=k_2+k_3$, this becomes
\begin{align}
     a_{I_1 I_2 I_3}&\,=\,\frac{1}{(k_1+1)(k_1+2)\,2^{k_1-1}} \langle C^{I_1} C^{I_2} C^{I_3} \rangle \;\equiv\; z(k_1) \langle C^{I_1} C^{I_2} C^{I_3} \rangle  \;.
\end{align}
Thus we have 
\be
    \langle C^{I_1} C^{I_2} C^{I_3} \rangle \,=\, \frac{a_{I_1 I_2 I_3} }{z(k_1)}\,, \qquad 
    \mathrm{for}~~~k_1 = k_2+k_3 \,.
\ee

Note that there are no contractions between $T_{I_2}$ and $T_{I_3}$ inside the extremal correlator. Therefore we can take the coincident limit $x_3\to x_2$ and obtain the two-point function 
\be
    \langle T_{I_1}(x_1) \left( T_{I_2} T_{I_3}\right)(x_2)  \rangle \,=\, \frac{1}{N} \frac{a_{I_1 I_2 I_3}}{z(k_2+k_3)}\frac{ \sqrt{(k_2+k_3) k_2 k_3}}{|x_1-x_2|^{2(k_2+k_3)} } \,.
\ee
From now on, we will suppress the coordinate dependence in most correlators.

We introduce the notation $\mathcal{C}^{I_1,I_2,I_3}$ for the mixing coefficients by writing the single-particle operator in the form 
\be
\label{eq:o-t-c}
    O_{I_1} \,=\, T_{I_1} - \sum_{I_2+I_3=I_1} \cC^{I_1,I_2,I_3}\, T_{I_2} T_{I_3} \,,
\ee
where the notation $I_2+I_3=I_1$ denotes that the sum is over all $I_2,I_3$ compatible with $k_2+k_3=k_1$ and charge conservation.
Note that a particular double-trace combination enters this sum either once if it is a square of a single trace, or twice if it is a product of two distinct operators. 

As discussed above, the mixing coefficients can be determined by requiring the vanishing of the two-point functions with all double-traces. 
One finds~\cite{Arutyunov:1999en}
\be
    \cC^{I_1,I_2,I_3} \,=\, \frac{\sqrt{(k_2+k_3) k_2 k_3}}{2N} 
    \langle C^{I_1} C^{I_2} C^{I_3} \rangle\,.
\ee

For our applications in the present work, we now specialize to single-particle operators which are SO(4) singlets, and also on those double-traces in such operators that are themselves SO(4) singlets, since only such terms will contribute to the expectation values in the backgrounds we study. Then the multi-index $I$ reduces to $I=(k,m)$.
We have
\be
    \cC^{(4,m),(2,n),(2,p)} \,=\, \frac{2}{N} \frac{a_{(4m)(2n)(2p)}}{z(4)} 
    \,, 
\ee
which agrees with the appropriate mixing coefficients in the single-particle operators presented in the previous subsection.
Thus \eqref{eq:o-t-c} becomes
\be
\label{eq:o-t-c-2}
   O_{4,m}  \,=\, T_{4,m} - \frac{2}{N} \frac{a_{(4m)(2n)(2p)}}{z(4)}  O_{2,n} O_{2,p} \,,
\ee
where we sum over all $n,p$ compatible with R-charge conservation.

It is convenient to express the precision holographic dictionary in terms of expectation values of operators with a different normalisation to \eqref{eq:normalizedTn}, \eqref{eq:normalized-On}, which arises from the supergravity computation of two and three-point functions~\cite{Skenderis:2006uy,Skenderis:2007yb}, that is,
\be
\label{eq:sugra-normalisations-1}
    \mathcal{T}_{k} \,=\,
    \mathcal{N}_k T_k\,, 
    \qquad\quad
     \mathcal{O}_{k} \,=\,
    \mathcal{N}_k O_k\,, 
\ee 
where 
\be
\label{eq:sugra-normalisations-2}
\mathcal{N}_2=\frac{N}{\pi^2} ~, \qquad\quad
\mathcal{N}_k = \frac{N}{\pi^2}(k-2)\sqrt{k-1} \qquad \mathrm{for} \quad k\neq2 ~.
\ee
We note that the same normalisation coefficient $\cN_k$ appears for both $\cT_k$ and $\cO_k$. 
Also, the normalization coefficients $\cN_k$ are the same for the SO(6) descendants as for the corresponding CPOs.

Finally, we express \eqref{eq:o-t-c-2} in explicit form for $\cT_{4,k}$:
\be \label{eq:tbc}
\cT_{4,m} \,=\, \cO_{4,m} + \frac{\cN_4}{\cN_2^2}\frac{2}{N}\frac{a_{(4m)(2n)(2p)}}{z(4)} \cO_{2,n} \cO_{2,p} \,,
\ee
where ${\cN_4}/{\cN_2^2} = {2\sqrt{3}\pi^2}/{N}$.

\subsection{Kaluza-Klein reduction to five dimensions}

We now review part of the Kaluza-Klein reduction of  type IIB supergravity on S$^5$ and its application to holography~\cite{Kim:1985ez,Lee:1998bxa,Skenderis:2006uy}.
We use $x$ to denote AdS$_5$ coordinates, $y$ to denote S$^5$ coordinates, and from now on we use $a,b$ to denote indices on S$^5$. 
We focus on the trace of the 10D metric on S$^5$, and the five-form with all legs on S$^5$. The fluctuations of these fields have KK expansions of the form
\begin{align}
\begin{aligned}
\label{eq:haa-f5-expand}
    \T{h}{a}{a} (x,y) &\,=\, \sum \pi_{I}(x) Y^{I}(y)\,, \cr
    f_{abcde} (x,y) &\,=\, \sum \Lambda^{I} b_{I}(x)  \epsilon_{abcde} Y^{I}(y)\,,
    \end{aligned}
\end{align}
where $Y^{I}$ denotes scalar spherical harmonics on S$^5$, see Appendix~\ref{app:conventions}. We focus on solutions that preserve SO(4) isometry in S$^5$, so as above, $I$ reduces to $I=(k,m)$.

From the fields $\pi_{I}$ and $b_{I}$, we form the combination $s_{I}$, which is part of the set of fields that diagonalize the equations of motion, and which is given by
\be
\label{eq:sk-defn}
    s_{I} \,=\, \frac{1}{20(k+2)} \left(\pi_{I}-10(k+4) b_{I}\right)\,.
\ee

One then defines fields $S_{I}$ which satisfy five-dimensional field equations that can be integrated into a five-dimensional action without derivative couplings.
The relation between the fields $S_I$ and $s_I$ takes the form~\cite{Lee:1998bxa,Skenderis:2006uy} 
\be
\label{eq:S-to-s}
    S_I \,=\, s_I+\sum_{J,K}\left(J^{IJK}s_J s_K+L^{IJK}D_{\mu}s_J D^{\mu}s_K \right) +O([s]^3)\,.
\ee
For the computation of expectation values of operators of dimension two, the linear term is sufficient. For the computation of expectation values of operators of dimension four, the linear and quadratic terms are sufficient~\cite{Skenderis:2006uy}.
We therefore write the relevant terms in the relations between the fields $S_{I}$ and the fields $s_{I}$ for $k=2,3,4$ as~\cite{Skenderis:2007yb}
\begin{align}
\label{eq:S2-S3-def}
    S_{I} &\,=\, w(s_{I}) s_{I}\,, \quad~~~ w(s_{I}) = \sqrt{\frac{8k(k-1)(k+2)z(k)}{k+1}}\,, \quad~~ k=2,3\,;\\
\label{eq:S-4-defn}
    S_{(4,m)} &\,=\, \frac{2\sqrt{3}}{5}\left(s_{(4,m)}-\frac{a_{(4m)(2n)(2p)}}{27 z(4)} \left(83 s_{(2,n)}s_{(2,p)}
    +7\mathcal{D}_{\mu}
   s_{(2,n)}\mathcal{D}^{\mu}s_{(2,p)} \right)\right)\,.
\end{align}

\subsection{Holographic dictionary in the single-trace basis}

We now clarify that the previous derivation of the precision holographic dictionary~\cite{Skenderis:2006uy,Skenderis:2007yb} was done in the trace basis. Using the notation $[S]_k$ for the coefficient of $z^k$ in the Fefferman-Graham expansion of the field $S$, the radial canonical momenta are given by~\cite{Skenderis:2006uy,Skenderis:2007yb}
\be
\label{eq:pis}
 \pi^{(2)}_{2,m} = 2 \left[S_{2,m}\right]_2
\,,\qquad~~~
\pi^{(2k-4)}_{k,m} \,=\, (2k-4) \left[S_{k,m}\right]_k \qquad (k\neq2)\;.
\ee

Firstly, for $\mathcal{O}_{2,m}=\mathcal{T}_{2,m}$ the holographic dictionary is given by\footnote{The relation between our notation and that of~\cite{Skenderis:2007yb} is $\mathcal{T}_{k} =
\mathcal{O}_{S_k}$.}
\be
\label{eq:cO2-sugra}
    \langle \mathcal{O}_{2,m} \rangle \,=\, \langle \mathcal{T}_{2,m} \rangle \,=\, \frac{N^2}{2\pi^2}  \pi^{(2)}_{2,m} \,.
\ee
Next, the combination of supergravity fields dual to the expectation value of the single-trace operator $\cT_{4,m}$ involves quadratic terms in $\pi^{(2)}_{2,m}$ as follows~\cite{Skenderis:2007yb}:
\be
\label{eq:cT-sugra}
    \langle \cT_{4,m} \rangle{} \,=\, \frac{N^2}{2\pi^2} \left( \pi^{(4)}_{4,m} + 
    \frac{2\sqrt{3}}{z(4)}
    a_{(4m)(2n)(2p)}
    \pi^{(2)}_{2,n} \pi^{(2)}_{2,p} \right)\,.
\ee
The quadratic terms were added in order to ensure consistency with the fact that in the trace basis, extremal three-point correlators are non-zero, and they factorize into a product of two-point functions~\cite{Skenderis:2006uy}.

\subsection{Holographic dictionary in the single-particle basis}

We now rewrite the holographic dictionary in the single-particle basis.
We propose that the holographic dictionary for the expectation value of the single-particle operator $\mathcal{O}_{4,m}$ takes the simple form
\be
\label{eq:cO-sugra}
    \langle \mathcal{O}_{4,m} \rangle \,=\, \frac{N^2}{2\pi^2}  \pi^{(4)}_{4,m} \,.
\ee

As a corollary of this proposal, we obtain the dictionary for the double-trace operators, as follows.  Combining \eqref{eq:cT-sugra} and \eqref{eq:cO-sugra} we obtain
\be
    \langle \cT_{4,m}\rangle \,=\,  \langle \cO_{4,m}\rangle + \frac{N^2}{2\pi^2} \frac{2\sqrt{3}}{z(4)} a_{(4m)(2n)(2p)} \pi_{2,n}^{(2)} \pi_{2,p}^{(2)} \,.
\ee
Combining this with  Eqs.~\eqref{eq:tbc} and \eqref{eq:cO2-sugra}, we obtain the following simple formula,
\be
\label{eq:dou-tr-dic}
   \langle\cO_{2,m} \cO_{2,n}\rangle \,=\, \left(\frac{N^2}{2\pi^2}\right)^2 \pi_{2,m}^{(2)} \pi_{2,n}^{(2)} \,=\, \langle\cO_{2,m}\rangle\langle\cO_{2,n}\rangle \,,
\ee
which is consistent with large $N$ factorization.

For use when working with ten-dimensional solutions, we now record formulae for the supergravity expectation values in terms of the fields $s_I$ that arise directly from the Fourier expansion of the 10D fields, 
Eq.\;\eqref{eq:sk-defn}. Firstly, combining~\eqref{eq:cO2-sugra} and \eqref{eq:S2-S3-def}, and using the square bracket notation introduced above \eqref{eq:pis}, we have
\begin{align}
    \label{eq:cO2-s}
    \langle \mathcal{O}_{2,m} \rangle  
    \,&=\,
    \frac{N^2}{2\pi^2} \frac{2\sqrt{8}}{3} \left[s_{(2,m)}\right]_2\;.
\end{align}

For $\cO_4$, we use the fact that for the following coefficient,
\be
    \pi^{(4)}_{4,m} \,=\, 4 \left[S_{4,m}\right]_4\,, 
\ee
there exists a form of the dictionary in terms of $s_{k,m}$ with no derivatives. Specifically, using~\cite{Skenderis:2007yb}
\be
\left[\mathcal{D}_{\mu}s_{(2,m)}\mathcal{D}^{\mu}s_{(2,n)}\right]_4 = 4 \left[s_{(2,m)} s_{(2,n)}\right]_4 \,,
\ee
from \eqref{eq:S-4-defn} one can derive
\be
    \left[S_{4,m}\right]_4
    \,=\, \frac{2\sqrt{3}}{5} \left[ s_{(4,m)}-\frac{37}{9 z(4)} a_{(4m)(2n)(2p)} s_{(2,n)} s_{(2,p)} \right]_4\,.   
\ee
Then using \eqref{eq:cO-sugra} and \eqref{eq:pis}, the holographic dictionary for $\mathcal{O}_{4,m}$ in terms of $s_I$ is given by
\be
\label{eq:O-4-s}
    \langle \mathcal{O}_{4,m} \rangle  
    \,=\,
    \frac{N^2}{2\pi^2} \frac{4\sqrt{3}}{5} \left[2s_{(4,m)}-\frac{74}{9 z(4)} a_{(4m)(2n)(2p)} s_{(2,n)} s_{(2,p)} \right]_4\,.
\ee
For comparison, we also record
the dictionary for $\mathcal{T}_{4,m}$~\cite[Eq.\;(2.27)]{Skenderis:2007yb},
\be
\label{eq:T-4-s}
    \langle  \mathcal{T}_{4,m} \rangle \,=\, \frac{N^2}{2\pi^2} \frac{4\sqrt{3}}{5} \left[2s_{(4,m)}+\frac{2}{3 z(4)} a_{(4m)(2n)(2p)} s_{(2,n)} s_{(2,p)} \right]_4 \,,
\ee
which is consistent with the discussion above. Finally, for the double-traces we have
\be
\label{eq:dou-tr-dic-s}
   \langle\cO_{2,m} \cO_{2,n}\rangle \,=\, \frac{8N^4}{9\pi^4}\left[s_{2,m}\right]_2\left[s_{2,n}\right]_2\,.
\ee


\section{Supergravity solutions}
\label{sec:backgd-fields}

\subsection{LLM solutions in AdS$_5\times$S$^5$}

We now review the asymptotically AdS$_5\times$S$^5$ LLM solutions~\cite{Lin:2004nb}. These solutions contain only the metric and five-form field strength. The solutions take the form
\begin{align}
     ds^2 &\,=\, -h^{-2}(dt+V_i dx^i)^2 + h^2(dy^2+dx^i dx^i) + y e^G d\Omega_3^2 +y e^{-G} d\medtilde{\Omega}_3^2 \;, \nonumber \\
    h^{-2} &\,=\, 2y \cosh{G}\,, \qquad~~ z=\frac{1}{2} \tanh{G}\,,\nonumber \\ 
    y\partial_y V_i &\,=\, \epsilon_{ij} \partial_j z \,, \qquad~
    y\left(\partial_i V_j - \partial_j V_i\right) = \epsilon_{ij} \partial_y z\,, \nonumber \\
   F_5 &\,=\, F_{\mu\nu} dx^{\mu} \wedge dx^{\nu} \wedge d\Omega_3 + \tilde{F}_{\mu\nu} dx^{\mu} \wedge dx^{\nu} \wedge d\medtilde{\Omega}_3\,, \nonumber \\
   F &\,=\, d B_t \wedge (dt+V) + B_t dV + d\hat{B}\,, \\
   \tilde{F} &\,=\, d \tilde{B}_t \wedge (dt+V) + \tilde{B}_t dV + d\tilde{B}\,, \nonumber \\
   B_t &\,=\, -\frac{1}{4} y^2 e^{2G}\,, \qquad~~~  \tilde{B}_t = -\frac{1}{4} y^2 e^{-2G}\,,\nonumber  \\
   d\hat{B} &\,=\,  -\frac{1}{4} y^3 *_3 d\left(\frac{z+\frac{1}{2}}{y^2}\right)\,, \qquad~~  d\tilde{B} =  -\frac{1}{4} y^3 *_3 d\left(\frac{z-\frac{1}{2}}{y^2}\right)\,,
   \nonumber 
\end{align}
where $*_3$ is the Hodge dual on $\mathbb{R}^3$ parameterized by $(y,x_1,x_2)$, in this section $i=1,2$, and where
\begin{align}
\begin{aligned}
   z(x_1,x_2,y) &\;=\; \frac{y^2}{\pi} \int_{R^2} \frac{z(x'_1,x'_2,0)dx'_1 dx'_2}{((x-x')^2+y^2)^2}\;,\\
   V_i(x_1,x_2,y) &\;=\; \frac{\epsilon_{ij}}{\pi} \int_{R^2} \frac{z(x'_1,x'_2,0) (x_j-x'_j) dx'_1 dx'_2}{((x-x')^2+y^2)^2} \;.
\end{aligned}
\end{align}
Regularity imposes a boundary condition on the $x_1$--$x_2$ plane at $y=0$, which ensures that one or both of the three-spheres in the metric shrink smoothly. The boundary condition is that the function $z(x_1,x_2,0)$ take the values $\pm 1/2$.
This can be depicted as a colouring of the $\mathbb{R}^2$ at $y=0$ into black and white regions. The  black regions correspond to droplets in a free fermion phase space, and their total area is fixed to be $2\pi$ in our conventions.
The free fermion description also plays a role in the holographic description of the corresponding CFT operators~\cite{Corley:2001zk,Berenstein:2004kk}. 

The two backgrounds we study in this paper both involve a single black region. 
The boundary of this region is specified by a function that we shall refer to as the  `profile function'. 
To write the profile functions, we introduce plane polar coordinates via $x_1=r\cos\tilde{\phi}$ and $x_2=r\sin\tilde{\phi}$.

The first profile corresponds to the configuration that lives in the consistent truncation. 
The small $\alpha$ expansion of the profile function is~\cite{Giusto:2024trt}\footnote{The relation to the notation of \cite{Giusto:2024trt} is $\alpha_{\mathrm{here}}=-\epsilon_{\mathrm{there}}$. We also note that in the very recent work~\cite{Aprile:2025hlt}, the parameter $\alpha$ is different to ours; it is given by $\alpha_{\mathrm{there}}
=
\frac{1}{\sqrt{2}}
\tanh
\left(
\frac{\epsilon_{\mathrm{there}}}{2}
\right)$ where the $\epsilon_{\mathrm{there}}$ of \cite{Giusto:2024trt} and \cite{Aprile:2025hlt} are the same up to an overall sign.\label{foot:2}
}
\begin{align}
\begin{aligned}
\label{eq:llm-SG-prof}
  r(\tilde\phi) \;=\; r_1(\tilde\phi) \,&=\;\! \sqrt{1+\alpha \cos(2 \tilde\phi)+\frac{\alpha^2}{2}\cos(4 \phi) + O(\alpha^3)} \; \\
\,&=\, 
  1+\frac{\alpha}{2}   \cos 2 \tilde\phi  + \frac{\alpha^2}{16}  \left(3\cos 4 \tilde\phi-1\right) + O(\alpha^3)\,.
\end{aligned}
\end{align}
In the second profile, studied in~\cite{Skenderis:2007yb,Turton:2024afd}, $r^2(\tilde\phi)$ involves a single mode:
\begin{align}
\begin{aligned}
\label{eq:llm-ripplon-prof}
r(\tilde\phi) \;=\; 
r_2(\tilde\phi) \,&=\;\! \sqrt{1+\alpha \cos(2 \tilde\phi)} \; \\
\,&=\, 
1+\frac{\alpha}{2}
\cos 2 \tilde\phi  -
\frac{\alpha^2}{16}  \left(\cos 4 \tilde\phi+1\right) + O(\alpha^3)\,,
\end{aligned}
\end{align}
where $\alpha < 1$ in the first line, and we expand for small $\alpha$ in the second line.
We see that the two profiles  \eqref{eq:llm-SG-prof},~\eqref{eq:llm-ripplon-prof} agree up to order $\alpha$ but differ at order $\alpha^2$. 
For small $\alpha$, each  profile describes a ripple on a unit circle.

We expand the metric and five-form field strength in small $\alpha$, up to order $\alpha^2$. For the metric, we introduce the notation
\be
\label{eq:g-0-2}
     g =g^{(0)} + \alpha \:\! g^{(1)} +\alpha^2 \:\! g^{(2)} \;.
\ee

When $\alpha=0$, both profiles reduce to the unit circle, which corresponds to empty global AdS. The quantities $z$ and $V$ are given in~\cite{Lin:2004nb}. 
After the change of coordinates
\be
y\,=\,R\cos\theta\,,\qquad r\,=\,\sqrt{R^2+1}\sin\theta\,,\qquad \tilde\phi \,=\, {\phi}-t\,,
\ee
the metric and flux are those of empty global AdS$_5\times$S$^5$, where the radii of AdS$_5$ and S$^5$ are both set to 1,
\begin{align}
\begin{aligned}
\label{eq:metric0}
ds^2 \,&=\,
-(R^2+1)dt^2 + \frac{dR^2}{R^2+1}+R^2 d\medtilde{\Omega}_3^2
+ d\theta^2 + \sin^2 \theta d\phi^2 + \cos^2\theta d\Omega_3^2 
\,, \cr
    F_5 \,&=\, R^3\, dt\wedge dR \wedge d\medtilde{\Omega}_3 + \cos^3\theta\sin\theta\, d\theta \wedge d\phi \wedge d{\Omega}_3\,.
\end{aligned}
\end{align}

\subsection{First-order backgrounds}

At first order in $\alpha$, the LLM solutions specified by both the profiles~\eqref{eq:llm-SG-prof} and~\eqref{eq:llm-ripplon-prof}
are the fields of a linearised supergraviton in global AdS$_5\times$S$^5$~\cite{Lin:2004nb,Rychkov:2005ji}. 
For later convenience we will 
generalise the discussion slightly, and present the first-order fields that correspond to the profile  
\be
\label{eq:prof-1-n}
r(\tilde\phi) \,=\;\! 1+\frac{\alpha}{2}
\cos(n \tilde\phi) \:\! 
+ O(\alpha^2) \;,
\ee
which for $n=2$ reduces to the linearisation of the profiles~\eqref{eq:llm-SG-prof},~\eqref{eq:llm-ripplon-prof}.

For holography, one must either fix a gauge, as done in~\cite{Kim:1985ez}, or work in a gauge-invariant formalism~\cite{Skenderis:2006uy}. We choose to impose de Donder-Lorentz gauge, in which the physical degrees of freedom are manifest, defined by
\be
    D^a h_{(ab)} \,=\, D^a h_{a\mu} \,=\, 0 \,.
\ee
Here round brackets on indices denote symmetric traceless part, $\mu,\nu,\ldots$ denote AdS$_5$ indices, and we remind the reader that  $a,b,\ldots$ denote S$^5$ indices.

The linearised diffeomorphism that converts the order $\alpha$ fields into de Donder-Lorentz gauge is given in~\cite{Grant:2005qc}. In this gauge, the first-order metric and four-form potential $A_4^{(1)}$ are given by
\begin{align}
\begin{aligned}
    \label{eq:g-1}
    g^{(1)}_{\mu\nu} &= \sum_{n=\pm 2} 
    \left(-\frac{6}{5} |n| \;\! \mathsf{s}_n \mathsf{Y}_n \, g^{(0)}_{\mu\nu} + \frac{4}{|n|+1} \mathsf{Y}_n \nabla_{(\mu} \nabla_{\nu)} \mathsf{s}_n \right), \qquad g^{(1)}_{\alpha\beta} = \sum_{n=\pm 2} 2|n| \;\! \mathsf{s}_n \mathsf{Y}_n \, g^{(0)}_{\alpha\beta}\,,
     \\
    &\qquad\qquad \qquad \qquad   A_4^{(1)} = \sum_{n=\pm 2} \left( \mathsf{Y}_n \star_{AdS_5} d\mathsf{s}_n - \mathsf{s}_n \star_{S^5} d\mathsf{Y}_n\right)\,,    
\end{aligned}    
\end{align}
where
\be
\label{eq:sn-Yn}
    \mathsf{s}_n \,=\, \frac{|n|+1}{8|n|(R^2+1)^{|n|/2}} e^{i n t}\,, \qquad \mathsf{Y}_n \,=\, e^{i n \phi} \sin^{|n|}\theta\,.
\ee

\subsection{Second-order backgrounds}

At order $\alpha^2$, we again impose de Donder-Lorentz gauge. The closed-form order $\alpha^2$ fields that follow from the single-mode profile given in Eq.~\eqref{eq:llm-ripplon-prof} were computed and converted into Kaluza-Klein form and de Donder-Lorentz gauge in~\cite{Turton:2024afd}. We do the same for the solution in the consistent truncation.
The order $\alpha^2$ fields are somewhat lengthy, and we shall not present them here. 
The explicit form of the order $\alpha^2$ metric of the solution specified by the single-mode profile can be found in~\cite{Turton:2024afd}.

Let us compare the two profile functions that we study.
The difference between the profiles in Eqs.~\eqref{eq:llm-SG-prof} and \eqref{eq:llm-ripplon-prof} is
\begin{align}
\label{eq:llm-prof-diff}
  r_2(\tilde\phi) -
  r_1(\tilde\phi) \;=\; 
  \frac{\beta}{2}  \cos 4 \tilde\phi \, + \:\! 
 O(\alpha^3)\,, \qquad \quad \beta \,=\, -\frac{\alpha^2}{2} \,.
\end{align}
where we have introduced $\beta$ by comparison with the linear terms in $\alpha$ in \eqref{eq:llm-SG-prof}, \eqref{eq:llm-ripplon-prof}.

Correspondingly, the difference between the two metrics is proportional to the metric of the linearized fields
\eqref{eq:g-1}--\eqref{eq:sn-Yn} for $n=4$, but with a coefficient proportional to $\alpha^2$ rather than $\alpha$.
Explicitly, denoting the second-order metric corresponding to the profile $r_1(\tilde{\phi})$ by $g_1^{(2)}$ and likewise for the second profile, we have
\be
\label{eq:prof-diff}
g_2^{(2)} - g_1^{(2)} \;=\; -\frac12 \;\! g^{(1)}_{(n=4)} \;.
\ee

We then expand the order $\alpha^2$ metric and five-form field strength in S$^5$ spherical harmonics. We focus on the components given in~\eqref{eq:haa-f5-expand}, and compute the fields $s_I$ defined in~\eqref{eq:sk-defn}, which we will use to compute holographic expectation values in the next section.

\section{Determining the dual CFT states}
\label{sec:evs-alpha-2}

\subsection{Ansatz for the CFT states}

In order to perform our precision holographic analysis, we now parameterise a family of CFT states that will include the two states of interest to us. 
The first ingredient will be the expansion of a coherent state built only from powers of $O_2$.
The second ingredient will be a term linear in $O_4$ at order $\alpha^2$, since we have seen that the difference between the Profile 1 and Profile 2 solutions is a single mode of frequency 4.

Several years ago, it was suggested that the solution corresponding to the single-mode profile may correspond to a coherent state built only from powers of $O_2$~\cite{Skenderis:2007yb}. 
However, more recently, it has been proposed that the solution in the consistent truncation is the one that is dual to a coherent state built only from powers of $O_2$~\cite{Giusto:2024trt}. 

We will find evidence in support of the more recent proposal, demonstrate that the CFT state dual to the solution defined by the single-mode profile contains $O_4$, and fix the coefficient of $O_4$ in the CFT state dual to that solution.

In preparation for constructing an appropriate Ansatz for the dual CFT states, we note that a solution composed of a set of linearised single-particle fluctuations corresponds to the following CFT state, to leading order in large $N$:
\be
\qquad\qquad\qquad\quad
    |\Phi\rangle \,=\, 
    |0\rangle + \sum\limits_n\delta_n\, O_n |0\rangle + O(\delta_n^2) 
    \,,\qquad \qquad 
    \delta_n \,=\, \frac{N\alpha_n}{2\sqrt{n}} \;,
\ee
where $\alpha_n$ is the amplitude of a linearised fluctuation of  frequency $n$, in the normalization used in Section~\ref{sec:backgd-fields}. The coefficient $\delta_n$ was worked out in~\cite{Skenderis:2007yb}.

We will proceed by making an Ansatz with arbitrary coefficients, $A,\,B,\,C$, that we will fix by holographic computations. These coefficients will be useful to illustrate the structure of the calculation. 
We have seen that the solutions corresponding to both Profile 1 and Profile 2 have, to linear order in $\alpha=\alpha_2$, a linearized mode of frequency 2. The difference between the profiles is a linearized mode of frequency 4, but with coefficient $\alpha_4=\beta=-\alpha^2/2$, see Eq.\;\eqref{eq:llm-prof-diff}. We thus consider the candidate set of states
\be
  \label{eq:state-ansatz}
    |H\rangle \,=\, \mathcal{N} \left( |0\rangle + C\;\! \delta_2\, O_2 |0\rangle + B\;\! \frac{\delta_2^2}{2}\,     \left(O_2\right)^2 |0\rangle + A \;\! \delta_4 \, O_4 |0\rangle + O(\delta_2^3,\delta_4^2) \right) \,,
\ee
where
\be
    \delta_2 \,=\, \frac{N\alpha}{2\sqrt{2}} \,, \qquad~~~
    \delta_4 \,=\,  -\frac{N\alpha^2}{8} \,.
\ee
To order $\alpha^2$, the norm of the state is $\mathcal{N} \;=\; 1-C^2\delta_2^2/2$. 

To fix the coefficients $A,B,C$, we first compute the relevant set of CFT expectation values up to order $\alpha^2$. These are protected quantities, and we compute them using the free theory. We continue to suppress the spacetime dependence of the correlators. 
We shall compute expectation values of the supergravity-normalized versions of the various operators, see Eqs.\;\eqref{eq:sugra-normalisations-1}--\eqref{eq:sugra-normalisations-2}.

Because of the orthogonality of the single-particle basis, the expectation values of the charged operators $\cO_2$, $\cO_4$, $\cO_2^2$ are simply proportional to the respective coefficients $C$, $A$, $B$, as follows. In the following equations, we write the leading-order result in the small $\alpha$ expansion (likewise for $\delta_{2}$, $\delta_{4}$); for ease of presentation, we suppress the notation $~\cdots + O(\alpha^{\#})$ etc. Firstly,
\be
\label{eq:cO2-CFT}
   \langle H|\mathcal{O}_2 |H\rangle \,=\, \mathcal{N}_2 C \delta_2 \langle  O_2^\dagger O_2 \rangle  \,=\, \frac{N^2}{2\sqrt{2}\pi^2} C \:\!\alpha \,.
\ee
Next, exploiting the orthogonality between single-particle and multi-particle operators, we find
\be
\label{eq:cO4-CFT}
    \langle H|\mathcal{O}_4|H\rangle \,=\, \mathcal{N}_4 A \delta_4 \langle  O_4^{\dagger} \, O_4 \rangle \,=\, -\frac{\sqrt{3}N^2}{4\pi^2} A \:\!\alpha^2\,.
\ee
For $\cO_2^2$, we obtain
\be
\label{eq:cO2sq-CFT}
    \langle H|\left( \mathcal{O}_2 \right)^2|H\rangle \,=\, \mathcal{N}_2^2 \frac{B \delta_2^2 }{2} \:\! \big\langle  \!\left(O_2^{\dagger}\right)^2 \left(O_2\right)^2 \big\rangle \,=\, \frac{N^4}{8\pi^4} B  \:\!\alpha^2 \,.
\ee

We now consider the neutral operators $\cO_{2,0}$ and $\cO_{4,0}$. At order $\delta_2^2$, and given the above Ansatz, these are only sensitive to the linear term in $\delta_2$, proportional to $O_2$, in~\eqref{eq:state-ansatz}. Firstly, at order $\delta_2^2$,
\begin{align}
\langle H|\cO_{4,0}|H\rangle \,=\,
\cN_4 C^2 \delta_2^2 \langle O_2^{\dagger} O_{4,0} O_2\rangle \,=\, 0 \,,
\end{align}
since $\langle O_2^{\dagger} O_{4,0} O_2\rangle $ is an extremal three-point correlator of single-particle operators. The CFT expectation value of $\cO_{2,0}$ at this order is~\cite{Skenderis:2007yb}
\begin{align}
    \label{eq:cO20-CFT}
    \langle H| \mathcal{O}_{2,0} |H\rangle \,=\, \cN_2 C^2 \delta_2^2 \langle O_2^{\dagger}O_{2,0}O_2\rangle  \,=\, \frac{N^2 \sqrt{2}}{4\sqrt{3}\pi^2} C^2 \:\! \alpha^2\,. 
\end{align}

\subsection{Analysis of the solution in the consistent truncation}

It has been argued that the LLM solution in the consistent truncation should be a coherent state composed of multi-traces of the operator $O_2$ only~\cite{Giusto:2024trt}.
If this is correct, then the single-particle operator $O_4$ should have zero expectation value in the corresponding heavy state. We will verify that this is indeed the case, and determine the state to order $\alpha^2$.

For Profile 1, using \eqref{eq:haa-f5-expand}--\eqref{eq:sk-defn}, the complete list of the fields $s_{I}$ up to order $\alpha^2$ is
\begin{align}
     s_{(2,2)} &\,=\,  \frac{3 e^{-2 i t}}{8 \left(R^2+1\right)}\alpha\,, \qquad s_{(2,-2)} \,=\, \frac{3 e^{2 i t}}{8 \left(R^2+1\right)}\alpha\,, \\
     s_{(2,0)} &\,=\,  \frac{\sqrt{3} \left(20 R^4+57 R^2+27\right)}{160 \left(R^2+1\right)^3}\alpha^2\,,\\
    s_{(4,0)} &\,=\, \frac{111 R^2+55}{192 \sqrt{5} \left(R^2+1\right)^3}\alpha^2 \,, \\
    s_{(4,4)} &\,=\, \frac{37 e^{-4 i t}}{64 \left(R^2+1\right)^2}\alpha^2  \,, \qquad s_{(4,-4)} \,=\,  \frac{37 e^{4 i t}}{64 \left(R^2+1\right)^2}\alpha^2 \,.   
\end{align}
To order $\alpha^2$ and exactly in $R$, we find the five-dimensional fields defined in Eq.~\eqref{eq:S-4-defn} to be
\begin{align}
     S_4 \,&=\, 0\,, \qquad
     S_{4,0} \,=\, 0\,.
\end{align}

To extract the supergravity expectation values, we must put the metric in Fefferman-Graham form. Recall that we denote the metric on empty global AdS$_5$ as $g_{\mu\nu}^0$. Similarly to~\cite{Skenderis:2007yb} we write the zero-mode on $S^5$ of the metric  as 
$g_{\mu\nu}^{(0)}+\tilde{h}^0_{\mu\nu}$ with $\tilde{h}^0_{\mu\nu}=h^0_{\mu\nu}+\frac{1}{3}\pi^0 g^{(0)}_{\mu\nu}$. 
Then to order $\alpha^2$, $g_{\mu\nu}^{(0)}+\tilde{h}^0_{\mu\nu}$ is given by
\begin{align}
    ds_{(0)}^2 \,=\, - dt^2\left(R^2 +1\right) & \left(1+\alpha^2 \frac{24 R^8+72 R^6+77 R^4+55 R^2+24}{32 \left(R^2+1\right)^4}\right) \nonumber\\ 
    &+ \frac{dR^2}{R^2+1}\left(1-\alpha^2 \frac{24 R^6+76 R^4+149 R^2+15}{32 \left(R^2+1\right)^4}\right) \\
    &+ R^2 d\Omega_3^2 \left(1-\alpha^2\frac{72 R^6+216 R^4+199 R^2+45}{96 \left(R^2+1\right)^3}\right)\,. \nonumber
\end{align}
The relation between the bulk radial coordinate and the Fefferman-Graham one is given by
\be
\label{eq:R-z-Prof-1}
    R \,=\, \frac{1}{z}-\frac{z}{4}-\frac{3\alpha^2}{16}z -\frac{\alpha^2}{64}z^3 +\ldots \,,
\ee
however only the first term  $R=\frac{1}{z}$ is relevant to our  analysis.

We now read off the supergravity expectation values using the holographic dictionary in Eqs.~\eqref{eq:cO2-s}--\eqref{eq:T-4-s}.
Using the Fefferman-Graham expansion and taking into account terms up to $\alpha^2$, the supergravity expectation values are
\begin{align}
\label{eq:evs-prof-1-a}
    \langle\mathcal{O}_{2}\rangle \,&=\, \frac{N^2}{2\sqrt{2}\pi ^2} e^{-2 i t}\alpha  \,, \qquad
    \langle\mathcal{O}_{2,0}\rangle \,=\, \frac{ N^2 \sqrt{2}}{4 \sqrt{3} \pi ^2 } \alpha^2 \,,
    \\ 
     \label{eq:evs-prof-1-b}
     \langle\mathcal{O}_{4}\rangle \,=\, 0 \,, 
      \qquad 
      \langle\mathcal{O}_{4,0}\rangle \,&=\, 0 \,, 
      \qquad 
       \langle\mathcal{T}_{4}\rangle \,=\, \frac{N^2  \sqrt{3}}{2 \pi ^2} e^{-4 i t}\alpha^2  \,,
       \qquad
      \langle\mathcal{T}_{4,0}\rangle \,=\,  \frac{N^2\sqrt{3}}{2\sqrt{5} \pi ^2}\alpha^2\,.
\end{align}
When comparing to CFT expectation values, we will set $t=0$.

Comparing the expectation values of the charged operators $\cO_2$, $\cO_2^2$ and $\cO_4$ to their respective CFT expectation values in Eqs.~\eqref{eq:cO2-CFT}--\eqref{eq:cO2sq-CFT}, we obtain to leading order in large $N$ 
\be
C \,=\, 1 \,, \qquad
B \,=\, 1 
, \qquad
A \,=\, 0 \,,
\ee
where we note that the double-trace factorization given in Eq.\;\eqref{eq:dou-tr-dic} implies $B=C^2$.

The expectation value of $\mathcal{O}_{4,0}$ at order $\alpha^2$ vanishes in supergravity, since $S_{4,0}=0$, and also in the CFT.  
The expectation value of $\cO_{2,0}$ gives an independent cross-check of $C=1$. The agreement of both $\langle\mathcal{O}_{4,0}\rangle$ and $\langle\mathcal{O}_{2,0}\rangle$ gives a non-trivial test of the completeness of the Ansatz~\eqref{eq:state-ansatz} to this order.

Thus, for the solution in the consistent truncation, we see that to order $\delta_2^2$ the following CFT state is consistent with all available expectation values of the known precision holographic dictionary:
\be
\label{eq:state_1}
    |H\rangle \,=\, \mathcal{N} \left( |0\rangle + \delta_2\, O_2 |0\rangle + \frac{1}{2}\delta_2^2 \left(O_2\right)^2 |0\rangle + O(\delta_2^3) \right)\,,  
\ee
where 
\be
\delta_2 \;=\; \frac{N\alpha}{2\sqrt{2}}\;,
\qquad\quad
\mathcal{N} \;=\; 1-\frac{\delta_2^2}{2} \;.
\ee
This is evidence in support of the proposal of~\cite{Giusto:2024trt} that the CFT state should be composed of multi-traces of $O_2$ only.\footnote{We note that a computation of $\langle\mathcal{O}_{4}\rangle \,=\, \langle\mathcal{O}_{4,0}\rangle \,=\, 0$ was reported in~\cite{Giusto:2024trt}, however the holographic dictionary quoted in Eq.\;(B.5) of that work is the dictionary for the single-trace operators, Eq.\;\eqref{eq:T-4-s}, rather than the dictionary for the single-particle operators, Eq.~\eqref{eq:O-4-s}.} It also agrees with the very recent refinement of this proposal in~\cite{Aprile:2025hlt}; the relation between the notations of these works is given in footnote \ref{foot:2}.

\subsection{Analysis of the single-mode ripple solution}

We now analyse the solution defined by the single-mode profile. 
For this solution, the fields $s_{2,\pm2}$, $s_{2,0}$ and $s_{4,0}$ are the same as for the solution in the consistent truncation. The only different component fields are $s_{4,\pm4}$, given below, and there are no other fields up to order $\alpha^2$. This reflects the fact that the  difference between solutions 1 and 2 is proportional to a single mode of frequency $\pm4$, see Eq.~\eqref{eq:llm-prof-diff}.  
For the solution specified by the single-mode profile, we have
\begin{align}
    s_{(4,4)} &\,=\, \frac{17 e^{-4 i t}}{64 \left(R^2+1\right)^2}  \alpha^2\,, \qquad s_{(4,-4)} \,=\,  \frac{17 e^{4 i t}}{64 \left(R^2+1\right)^2}\alpha^2 \,.
\end{align}
We therefore find the five-dimensional fields $S_{4,m}$ to be
\begin{align}
     S_4 &\,=\, -\frac{\sqrt{3} \, e^{-4 i t}}{8 \left(R^2+1\right)^2} \alpha^2 \,, \qquad
     S_{4,0} \,=\,  0\,.
\end{align}
The Fefferman-Graham expansion depends only on the zero-modes of the ten-dimensional metric reduced on S$^5$. Therefore it is the same as that of Profile 1 in Eq.~\eqref{eq:R-z-Prof-1}.

Most of the supergravity expectation values for Profile 2 are the same as those for Profile 1, and are not sensitive to the difference between the two profiles.

However, importantly, for the single-mode profile, the supergravity expectation values of ${\cO}_{4}$ and $\cT_4$ differ from the Profile 1 expressions in Eqs.\;\eqref{eq:evs-prof-1-a}, \eqref{eq:evs-prof-1-b}. The full set of non-zero supergravity expectation values up to order $\alpha^2$ for the single-mode profile background are:
\begin{align}
\begin{aligned}
\label{eq:evs-prof-2-a}
    \langle\mathcal{O}_{2}\rangle \,&=\, \frac{N^2}{2\sqrt{2}\pi ^2} e^{-2 i t}\alpha  \,, \qquad
    \langle\mathcal{O}_{2,0}\rangle \,=\, \frac{ N^2 \sqrt{2}}{4 \sqrt{3} \pi ^2 } \alpha^2 \,,
     \qquad 
      \langle\mathcal{O}_{4,0}\rangle \,=\, 0 \,, 
    \cr 
     \langle\mathcal{O}_{4}\rangle \,&=\, -\frac{\sqrt{3} N^2}{4 \pi ^2}  e^{-4 i t} \alpha^2 \,, 
          \qquad 
       \langle\mathcal{T}_{4}\rangle \,=\, \frac{N^2  \sqrt{3}}{4 \pi ^2} e^{-4 i t}\alpha^2  \,,
       \qquad
      \langle\mathcal{T}_{4,0}\rangle \,=\,  \frac{N^2\sqrt{3}}{2\sqrt{5} \pi ^2}\alpha^2\,,
    \end{aligned}
\end{align}

Comparing the supergravity and CFT expressions for the expectation values of $\cO_2$, $\cO_2^2$, and particularly $\cO_4$, we find that for the solution defined by the single-mode profile, 
the dual CFT state has coefficients
\be
\label{eq:abc-prof-2}
C \,=\, 1 \,, \qquad
B \,=\, 1 
, \qquad
A \,=\, 1 \,.
\ee
Thus, for the solution defined by the single-mode profile, to leading order in large $N$ and to order $\alpha^2$, the following CFT state is consistent with all available expectation values of the known precision holographic dictionary: 
\be
\label{eq:state_2}
       |H\rangle \,=\, \mathcal{N} \left( |0\rangle +  \delta_2\, O_2 |0\rangle + \frac{\delta_2^2}{2}\,     \left(O_2\right)^2 |0\rangle +  \delta_4 \, O_4 |0\rangle + O(\delta_2^3,\delta_4^2) \right) \,,
\ee
where
\be
    \delta_2 \,=\, \frac{N\alpha}{2\sqrt{2}} \,, \qquad~~~
    \delta_4 \,=\,  -\frac{N\alpha^2}{8} \,, \qquad~~~
\mathcal{N} \;=\; 1-\frac{\delta_2^2}{2} \,.
\ee

We note that in the holographic study of the single-mode profile solution in~\cite{Skenderis:2007yb}, the single-trace basis was used. The expectation values of $\cO_2$, $\cO_{2,0}$, and $\cT_{4,0}$ in \eqref{eq:evs-prof-2-a} agree with those reported in~\cite{Skenderis:2007yb}. 
By contrast, the expectation value of $\cT_4$ in this background was not examined in that work.
By computing the expectation value of $\cO_4$, we have demonstrated the presence of $O_4$, and fixed its coefficient in the dual CFT state.

\section{Cross-check at cubic order}
\label{sec:evs-alpha-3}

In this section we consider the solution defined by the single-mode profile, and make a cross-check of the dual CFT state given in Eq.\;\eqref{eq:state_2}. We do so by considering the charged R-symmetry descendant operator $\cO_{4,2}$. 
For this operator, the CFT state \eqref{eq:state_2} up to order $\alpha^2$ gives rise to an expectation value at order $\alpha^3$. We demonstrate that this is precisely reproduced by the corresponding supergravity analysis.

To do so, we compute the closed-form LLM solution that follows from the single-mode profile to order $\alpha^3$, and bring it to de Donder-Lorentz gauge. We then extract the KK fields $s_I$ as before. The complete list of the fields $s_I$ at order $\alpha^3$ is
\begin{align}
     s_{(4,2)} &\,=\, \frac{e^{-2 i t}\left(119 R^6 + 459 R^4 + 477 R^2 + 116 \right)}{224 \sqrt{10} \left(R^2+1\right)^5}  \alpha^3 \,,\\
     s_{(2,2)} &\,=\, \frac{e^{-2 i t}\left(720 R^8 + 2112 R^6 + 2247 R^4 + 836 R^2 + 233\right)}{2560\left(R^2+1\right)^5} \alpha^3\,.  
\end{align}

We note in passing that the order $\alpha^3$ term in $s_{(2,2)}$ implies an $\alpha^3$ correction to the supergravity expectation value of $\langle  \mathcal{O}_{2} \rangle$, which in our parameterization takes the form
\be
    \langle  \mathcal{O}_{2} \rangle \,=\, 
    \frac{N^2}{2\sqrt{2}\pi ^2} \Big( \alpha + \frac{3}{4}\alpha^3 \Big) \,.   
\ee
This would be sensitive to a possible term proportional to $\alpha^3\ket{O_2}$ in the dual state, however we have expanded the dual state only to order $\alpha^2$, so this is beyond the precision to which we work. We thus focus instead on $\cO_{4,2}$.

To compute the supergravity expectation value of $\cO_{4,2}$, we compute the 5D field $S_{4,2}$. 
Using Eq.~(\ref{eq:S-4-defn}) we obtain 
\be
    S_{4,2} \,=\, -\sqrt{\frac{3}{10}} \frac{\left(2800 R^6+10567 R^4+8296 R^2+2909\right) e^{-2 i t}}{11200 \left(R^2+1\right)^5} \alpha^3\,.
\ee
The relevant coefficient in the Fefferman-Graham expansion is
\be
    \left[S_{4,2}\right]_4 \,=\, -\frac{1}{4} \sqrt{\frac{3}{10}} e^{-2 i t} \alpha^3 \,,
\ee
and so the holographic expectation value is 
\be
\label{eq:h-ev-O42}
    \langle \mathcal{O}_{4,2} \rangle \,=\, -\frac{N^2}{2 \pi ^2}\sqrt{\frac{3}{10}} e^{-2 i t} \alpha^3 \,.
\ee

We now compare this result to the CFT, i.e.~we compute $\langle \mathcal{O}_{4,2} \rangle$ in the state~\eqref{eq:state_2}:
\be
\label{eq:o42-cft}
    \langle H| \mathcal{O}_{4,2} |H\rangle \,=\, \mathcal{N}_4 
    \delta_2 \left(\frac{\delta_2^2}{2}\big\langle \big( O_2^\dagger \big)^2 \:\! O_{4,2} \:\! O_2 \big\rangle+\delta_4\langle O_4^\dagger \:\! O_{4,2} \:\! O_2 \rangle \right) \,.
\ee
Firstly,  $\big\langle (O_2^\dagger)^2 \:\! O_{4,2} \:\! O_2 \big\rangle$ gives a subleading contribution in large $N$ compared to the term proportional to $\langle O_4^\dagger \:\! O_{4,2} \:\! O_2 \rangle$, see Appendix~\ref{app:sec-5-details} for details. 

Next, we consider $\langle O_4^\dagger \:\! O_{4,2} \:\! O_2 \rangle$. Expanding $O_4^\dagger$ in terms of single-traces, this is 
\be
\label{eq:ev-1}
    \langle O_4^\dagger O_{4,2} O_2 \rangle \,=\, \langle T_4^\dagger O_{4,2} T_2 \rangle -\frac{2}{N} \big\langle \;\! \big(O_2^\dagger\big)^2 O_{4,2} \:\! O_2 \big\rangle\,.
\ee
We have determined already that the second correlator on the right-hand side of Eq.\;\eqref{eq:ev-1} is subleading compared to the first one. The first term evaluates to
\be
\label{eq:a-3-int}
    \langle T_4^\dagger O_{4,2} T_2 \rangle \,=\, \langle T_4^\dagger T_{4,2} T_2 \rangle - \frac{4\sqrt{3}}{\sqrt{10}N} \langle T_4^\dagger \left(T_{2,0} T_2\right) T_2 \rangle\,. 
\ee
Furthermore, the second term on the right-hand side  of Eq.\;\eqref{eq:a-3-int} is suppressed at large $N$ compared to the first term, see Appendix~\ref{app:sec-5-details} for details. So, to leading order in large $N$, we have
\be
\label{eq:a-3-int-2}
    \langle O_4^\dagger \:\! O_{4,2} \:\! O_2 \rangle \,=\, \langle T_4^\dagger \:\! T_{4,2} \:\! T_2 \rangle \,=\, \frac{4}{\sqrt{5}N} \,.
\ee
The CFT expectation value at this order, from \eqref{eq:o42-cft}, is then
\be
   \langle H| \mathcal{O}_{4,2} |H\rangle \,=\, \mathcal{N}_4 \frac{4 \:\! \delta_2 \:\! \delta_4}{\sqrt{5}N} \,=\, - \frac{N^2}{2\pi^2} \sqrt{\frac{3}{10}} \alpha^3\,, 
\ee
which precisely agrees with the holographic result~\eqref{eq:h-ev-O42}. 
This is a non-trivial cross-check of the CFT state dual to the single-mode profile solution given in Eq.\;\eqref{eq:state_2}.
This also represents a non-trivial check of the proposed holographic dictionary for single-particle operators in Eqs.~\eqref{eq:cO-sugra}, \eqref{eq:O-4-s}.


\section{Discussion}
\label{sec:disc}

In this paper we revisited the AdS$_5$/CFT$_4$ precision holographic dictionary for heavy-light three-point correlators. We clarified that it was originally expressed in the single-trace basis, and rewrote it in the single-particle basis. 
The holographic dictionary takes a simpler form in the single-particle basis, see Eq.~\eqref{eq:cO-sugra}.

The single-particle basis gave a distinct advantage over the trace basis for our computation, because the dual CFT states we studied involve both single and double-trace operators. 
The orthogonality of the single-particle basis meant that each coefficient in our Ansatz for the dual CFT states was controlled by a single expectation value, see Eqs.\;\eqref{eq:state-ansatz}--\eqref{eq:cO2sq-CFT}.

We performed a holographic analysis of the two LLM supergravity solutions under consideration, perturbatively in $\alpha$. From the asymptotic expansion of the appropriate gauge-invariant fields, we first computed the expectation values of the operators that directly control the coefficients in the Ansatz for the dual CFT states. These determine the dual CFT states up to order $\alpha^2$ and at leading order in large $N$, see Eqs.~\eqref{eq:state_1} and \eqref{eq:state_2}.

We also computed the supergravity expectation values of a set of R-symmetry descendants of chiral primaries. All of these resulted in precise agreement between gravity and CFT. 
We computed all expectation values that arise up to order $\alpha^2$, and also the expectation value of the R-symmetry descendant $O_{4,2}$ in the solution defined by the single-mode profile, which arises at order $\alpha^3$.
The agreement of these expectation values constitutes a set of non-trivial cross-checks of both the dual CFT states and the precision holographic dictionary in the single-particle basis.

Our results represent evidence in favour of the proposal that the dual CFT state of the LLM solution that lies in the consistent truncation is a coherent state composed only of powers of the dimension-two chiral primary $O_2$~\cite{Giusto:2024trt,Aprile:2025hlt}. 
We showed that the solution defined by a single-mode profile contains, by contrast, the dimension-four chiral primary $O_4$ at order $\alpha^2$, and determined the coefficient of $O_4$ at this order.

This raises a natural question for future work. That is, for the solution defined by the single-mode profile, what the dual CFT state is at higher orders in $\alpha$. 
It is natural to expect that a sequence of higher-dimension operators appears at successive orders in $\alpha$. 
Such terms can in principle be analyzed by extending the precision holographic dictionary to operators with dimensions higher than four, which has not been done to date.

To extend the holographic dictionary to operators of dimension six, it would first be necessary to solve explicitly the relation between the five-dimensional fields $S_I$ and the fields $s_J$, see Eq.\;\eqref{eq:S-to-s}, up to cubic order in $s_J$, to account for cubic terms in $s_{2,m}$. This entails expanding the equations of motion up to cubic order in fluctuations and performing field redefinitions to remove derivative couplings~\cite{Lee:1998bxa,Skenderis:2006uy}. 
This would enable a study of the dual CFT state of the solution defined by a single-mode profile up to order $\alpha^3$.

More broadly, the LLM family of solutions is a large class, and there are even larger classes of 1/4 and 1/8-BPS solutions, see e.g.~\cite{Chen:2007du,Lunin:2008tf,Jia:2023iaa}. 
It would be interesting to perform precision holographic analyses of more general asymptotically AdS$_5 \times$S$^5$ solutions.
We re-emphasize that the holographic dictionary in the single-particle basis gives an advantage over the single-trace basis, and we expect that this form of the dictionary will prove useful for future studies.

\vspace{3mm}

\section*{Acknowledgements}

For valuable discussions, we thank Stefano Giusto, Rodolfo Russo, and Kostas Skenderis. The work of D.T.~was supported by a Royal Society Tata University Research Fellowship. The work of A.T.~was supported by a Royal Society Research Fellows Enhanced Research Expenses grant.


\vspace{2mm}

\begin{appendix}

\section{Spherical harmonics and symmetric traceless tensors}
\label{app:conventions}

We consider the following single-trace operators,
\begin{align}
\widehat{T}_{I} \,&=\, C^I_{i_1 \cdots i_k} 
\tr (\Phi^{i_1}\cdots \Phi^{i_k})
\end{align}
where $C^I$ is a totally symmetric traceless rank-$k$ tensor of SO(6), which live in half-BPS multiplets~\cite{Lee:1998bxa}. 
The tensors $C^I$ are unit-normalized, 
$\langle C^I C^J \rangle \equiv  C^{I}_{i_1 \cdots i_k} C^{J}_{i_1 \cdots i_k}=\delta^{IJ}$. Each $C^I$ corresponds to a scalar SO(6) spherical harmonic via $Y^I=C^I_{i_1 \cdots i_k} x^{i_1}\cdots x^{i_k}$ for unit-norm vectors $x^{i_j} \in \mathbb{R}^6$.

Scalar spherical harmonics on S$^5$ satisfy 
\be
\square_{\mathrm{S}^5} Y^{I} \;=\; 
\Lambda^{I} Y^{I} \;,
\qquad
\Lambda^{I} \;=\; -k(k+4),
\qquad
k=0,1,2,\ldots
\ee
We write the metric on S$^5$ as
\be
ds_{\mathrm{S}^5}^2 \,=\, d\theta^2 + \sin^2 \theta d\phi^2 + \cos^2\theta d\Omega_3^2 \;.
\ee
In this paper we restrict to harmonics with SO(4) isometry, which depend only on $\theta$ and $\phi$. Then the multi-index $I$ reduces to $I=(k,m)$ and the scalar harmonics $Y^{(k,m)}$ are given in terms of hypergeometric functions~\cite{Skenderis:2006uy}.

Denoting the area of the unit five-sphere by $\omega_5=\pi^3$, the scalar spherical harmonics are then normalized as
\begin{align}
\frac{1}{\omega_5}\int_{S^5} Y^{I_1} Y^{I_2} \;=\; z(k) \delta^{I_1 I_2} 
\,, \qquad
z(k) \;\equiv\; \frac{1}{2^{k-1} (k+1)(k+2)}\,.
\end{align}

The triple intersection constants $a_{I_1I_2I_3}$ are defined by
\be 
\label{eq:trip}
a_{I_1 I_2 I_3} \;\equiv\; \frac{1}{\omega_5}\int_{S^5} Y^{I_1} Y^{I_2} Y^{I_3} \,.
\ee
In terms of the SO(6) invariant (see e.g.~\cite{Arutyunov:2018neq})
\begin{align}
\label{eq:trip-c}
\left\langle C^{I_1} C^{I_2} C^{I_3}  \right\rangle \,\equiv\, 
C^{I_1}_{i_1 \ldots \;\! i_{\alpha_2} \:\! j_{1} \:\! \ldots \:\! j_{\alpha_3}} C^{I_2}_{j_1\ldots \:\! j_{\alpha_3} \:\! l_{1} \ldots \:\! l_{\alpha_1}}C^{I_3}_{l_1\ldots l_{\alpha_1} \:\! i_{1} \ldots \:\! i_{\alpha_2}}\,,
\end{align}
we have the following relation, used in the main text around Eq.\;\eqref{eq:a-c-main},
\be
     a_{I_1 I_2 I_3}\,=\, \frac{1}{\left(\frac{1}{2}\Sigma+2\right)! \, 2^{\frac{1}{2}\left(\Sigma-2\right)}}\frac{k_1! k_2! k_3!}{\alpha_1! \alpha_2!\alpha_3!}\langle C^{I_1} C^{I_2} C^{I_3} \rangle \,,
\ee
where $\Sigma=k_1+k_2+k_3$, $\alpha_1=\frac12(k_2+k_3-k_1)$, and similarly for $\alpha_2,\alpha_3$.

We record here some useful values of triple intersection constants:  
\begin{align}
\begin{aligned}
\label{eq:triple-overlaps}
    a_{(4,4)(2,-2)(2,-2)} \,&=\,z(4) \,, \qquad~~~ 
    a_{(4,0)(2,0)(2,0)} \,=\, \frac{3 z(4)}{2 \sqrt{5}} \,,
   \cr 
   a_{(4,0)(2,2)(2,-2)} \,&=\, \frac{z(4)}{2 \sqrt{5}}\,, \qquad a_{(4,2)(2,0)(2,-2)}  \,=\, z(4)\sqrt{\frac{3}{10}}\,.  
\end{aligned}
\end{align}


\section{CFT expectation values at cubic order}

\label{app:sec-5-details}

In this appendix we record some details of the free CFT computations in Section~\ref{sec:evs-alpha-3}. We first describe the computation of the expectation value of $\mathcal{O}_{4,2}$.
Our starting point is Eq.~\eqref{eq:o42-cft}, i.e.
\be
\label{eq:o42-cft-app}
    \langle H| \mathcal{O}_{4,2} |H\rangle \,=\, \mathcal{N}_4  
    \delta_2 \left(\frac{\delta_2^2}{2}\big\langle (O_2^\dagger)^2 O_{4,2} O_2 \big\rangle+\delta_4\langle O_4^\dagger O_{4,2} O_2 \rangle \right) \,.
\ee
Firstly, we verify explicitly that $\big\langle (O_2^\dagger)^2 O_{4,2} O_2 \big\rangle$ gives a subleading contribution in large $N$ compared to the term proportional to $\langle O_4^\dagger O_{4,2} O_2 \rangle$.
Rewriting in the trace basis, we have
\be
    \langle (O_2^\dagger)^2 O_{4,2} O_2 \rangle \,=\, \langle (T_2^\dagger)^2 T_{4,2} T_2 \rangle - \frac{4\sqrt{3}}{\sqrt{10}N} \langle (T_2^\dagger)^2 \left(T_{2,0}T_2\right) T_2 \rangle \,.
\ee
Explicitly, the relevant terms that contribute to these correlators are
\begin{align}
    \langle (T_2^\dagger)^2 T_{4,2} T_2 \rangle &\,=\,\frac{1}{2\sqrt{5}N^5}\langle \left(\mathrm{Tr}\bar{Z}^2\right)^2 \mathrm{Tr}Z^3 \bar{Z} \, \mathrm{Tr} Z^2 \rangle\,,\\
    \langle (T_2^\dagger)^2 \left(T_{2,0}T_2\right) T_2 \rangle &\,=\, \frac{\sqrt{2}}{4\sqrt{3}N^5}  \langle \left(\mathrm{Tr}\bar{Z}^2\right)^2 \left(\mathrm{Tr}Z \bar{Z} \, \mathrm{Tr} Z^2 \right) \mathrm{Tr} Z^2 \rangle\,.
\end{align}
Free-field Wick contractions give
\begin{align}
    \langle (T_2^\dagger)^2 T_{4,2} T_2 \rangle &\,=\, \frac{16}{\sqrt{5}N^2} +
    O(N^{-4})\,,\\
    \langle (T_2^\dagger)^2 \left(T_{2,0}T_2\right) T_2 \rangle &\,=\, \frac{4\sqrt{2}}{\sqrt{3}N} +
    O(N^{-4})\,.
\end{align}
Therefore 
\be
    \langle (O_2^\dagger)^2 O_{4,2} O_2 \rangle \,=\, O(N^{-4})\,,
\ee
and so
\be    \mathcal{N}_4\delta^3\langle (O_2^\dagger)^2 O_{4,2} O_2 \rangle \,=\, \alpha^3 \:\! O( N^0)\,.
\ee

Now we turn to $\langle O_4^\dagger O_{4,2} O_2 \rangle$. The structure of this computation is described around Eq.\;\eqref{eq:a-3-int} of the main text; here we describe in more detail the fact that the following correlator gives a subleading contribution at large $N$:
\be
    \langle T_4^\dagger \left(T_{2,0} T_2\right) T_2 \rangle \,=\, \frac{\sqrt{2}}{4\sqrt{3}N^5} \langle \mathrm{Tr}\bar{Z}^4 \left(\mathrm{Tr}Z\bar{Z} \mathrm{Tr}Z^2\right) \mathrm{Tr}Z^2\rangle\,.
\ee
Free-field Wick contractions give
\be
    \langle \mathrm{Tr}\bar{Z}^4 \left(\mathrm{Tr}Z\bar{Z} \mathrm{Tr}Z^2\right) \mathrm{Tr}Z^2\rangle \,=\, 8N^3+
    O(N)\,,
\ee
so we find
\be
\frac{1}{N}
    \langle T_4^\dagger \left(T_{2,0} T_2\right) T_2 \rangle \,=\, \frac{2\sqrt{2}}{\sqrt{3}N^3}+ 
    O(N^{-5})\,.  
\ee
Thus, to leading order in large $N$, we have 
\be
    \langle O_4^\dagger O_{4,2} O_2 \rangle \,=\, \langle T_4^\dagger T_{4,2} T_2 \rangle +O(N^{-3})\,,
\ee
leading to Eq.\;\eqref{eq:a-3-int-2} of the main text.

\end{appendix}

\newpage

\begin{adjustwidth}{-3mm}{-3mm} 
\bibliographystyle{utphys}      
\bibliography{microstates}       

\end{adjustwidth}


\end{document}